\documentstyle[sprocl]{article}

\def\Journal#1#2#3#4{{#1} {\bf #2}, #3 (#4)}

\def\AJ{\em AJ}
\def\APJ{\em ApJ}

\def\be{\begin{equation}}
\def\ee{\end{equation}}
\def\bea{\begin{eqnarray}}
\def\eea{\end{eqnarray}}
\def\kms{{\rm km}\,{\rm s}^{-1}}
\def\dol{{D_{\rm ol}}}
\def\dls{{D_{\rm ls}}}
\def\dos{{D_{\rm os}}}

\begin{document}

\title{MICROLENSING: PROSPECTS FOR THE FUTURE}

\author{ANDREW GOULD\footnote{Alfred P.\ Sloan Foundation Fellow}}

\address{Dept of Astronomy, 174 W 18th Ave, Columbus, OH 43210, USA}


\maketitle\abstract{
Four ongoing microlensing experiments have produced important new results but
also big puzzles, the major one being that the expected classes of lenses
cannot account for the observed distribution of time scales.  I discuss future
experiments that could resolve these puzzles.  By far the most important
would be to launch a parallax satellite into solar orbit.  I also discuss
a number of non-dark-matter applications of microlensing, including searching
for planets, measuring the rotation speeds of giant stars, and imaging a black
hole at the center of a quasar.
}
  
\section{Introduction}

Within a few short years, microlensing has been transformed from an apparently 
hair-brained idea of a theorist~\cite{Pac} into a practical experimental 
approach to search for dark matter.\cite{machoI}$^{\!-\,}$\cite{duo}  
What are the most pressing questions posed
by the current observations?  Are there practical approaches for resolving
these questions?  How can we expect microlensing to develop as a technique
in the future?

\section{Major Results, Major Puzzles}

	The latest results from the MACHO collaboration's observations toward
the Large Magellanic Cloud (LMC) reported here by Sutherland seem to imply that
of order half of the dark matter between the Sun and the LMC is composed of
compact objects.\cite{machoII}  However, the best estimate of the mean
mass of these objects is $\sim 0.4\,M_\odot$.  If such objects were composed
of hydrogen, they would be M or K dwarf stars and would certainly have been
noticed (as I reported here earlier~\cite{fgb}).  This mass range is consistent
with white dwarfs, but the production of white dwarfs would be accompanied
by the generation of large amounts of carbon and other heavy elements,
contrary to observations.  In addition, we are now able to directly search
for halo white dwarfs and do not find them, although the limits are not yet
very strong.\cite{fgb}  One possibility is that the objects being detected
are not actually in the halo, but rather are in the disk of our Galaxy or
in the LMC itself.\cite{sahu}  While there are theoretical arguments against
major contributions from both the disk~\cite{gbfI} and the LMC~\cite{gouldLMC},
it is nonetheless striking that of the 8 events reported by MACHO, one 
(the binary event) is very likely in the LMC and another is quite possibly in 
the disk.  The proper motion (angular speed) of the binary is apparently 
$20\,\kms\,D_{\rm LMC}^{-1}$, where $D_{\rm LMC}\sim 50\,$kpc is the distance
to the LMC.  This would be quite consistent with an LMC object, but halo 
objects should be moving 50 times faster.  The unlensed light in MACHO event
5 has a color and flux consistent with a foreground disk M dwarf, but not
consistent with any known population of LMC stars.  It is natural to suppose
that this unlensed light is in fact the lens.  

	It is also possible that the masses of the lenses are being 
overestimated.  The mass estimates are based on the measured timescales
of the events and on a model for the physical and velocity distributions of the
lenses.  Models are required because the timescale, $t_e$, is related to the
underlying physical parameters of interest in a complicated way.  The
physical size of the Einstein ring, $r_e$, is given by
\be
r_e = \sqrt{4 G M \dol\dls\over c^2 \dos},\label{eq:redef}
\ee
where $M$ is the lens mass and $\dol$, $\dos$, and $\dls$ are the distances
between the observer, lens, and source.  Then,
\be
t_e = {r_e\over |{\bf v}|}, \label{eq:tedef}
\ee
where $\bf v$ is the transverse velocity of the lens relative to the line
of sight from the observer to the source.  If the lenses are in the halo,
their space velocities must be of order $\sim 200\,\kms$ by the virial theorem
and from measurements of the Galactic potential.  However, the exact structure
of the potential is not known.  Moreover, the flattening of the halo is not
known and this could affect the timescale through the ``$\dol$'' term in
Eq.~\ref{eq:redef}.  Furthermore, if the halo were rotating, the lenses might
have large absolute space motions but their projected velocities $\bf v$ might
nonetheless be small depending on the relative orientation of the lens
and LMC velocities.  In brief, the interpretation of the lensing observations
toward the LMC is far from unique.

	The observations toward the Galactic bulge are even more puzzling.
Although many seem to believe that the current observations are well
explained as lensing by ``known'' populations of stars, the required 
populations are not at all ``known'' by the people who study them.  On the
contrary, the events seem to require an enormous population of ``unknown''
brown dwarfs. What {\it is} known, is that half the dynamical mass of the bulge
is accounted for by bulge stars with masses $M>0.5\,M_\odot$ that can be
seen using the Hubble Space Telescope (HST).\cite{lbh}  (See also my other 
contribution to these proceedings.)\ \  These known stars contribute only a
tiny fraction of the observed microlensing events and almost none of the
short ones.\cite{han}  Even if the bulge mass function is extended to the
brown dwarf limit using the locally measured disk mass function, very few of
the short events are accounted for.  Only if the remaining 1/3 of the bulge 
mass is put in brown dwarfs $(M\sim 0.08\,M_\odot$), can the short events be
explained.\cite{han}  Of course, the problems of interpretation are similar
to those for the LMC.  While the kinematics of luminous stars can be directly
measured, those of the unobserved lenses (the great majority) cannot.  One
can assume that the kinematics are similar and this assumption may be a
reasonable one.  However, if the inference is the existence of a huge a
population of otherwise unobservable objects, one must question whether their
kinematics can really be assumed to be the same.

	One would like to actually measure both the kinematics and the masses
of the individual lenses toward both the LMC and the bulge, rather than relying
on statistical arguments.

\section{Parallax Satellite}

	By far the best idea for learning more about the lenses is to launch
a parallax satellite into solar orbit.\cite{refsdal}$^{\!-\,}$\cite{gouldsatI}
For objects in the relevant mass
range, $0.1\,M_\odot<M<1\,M_\odot$, the Einstein ring given by 
Eq.~\ref{eq:redef}
is $r_e\sim 1$--5 astronomical units (AU).  Hence, if one were to observe
the event from $\sim 1$ AU away, the event would look very different.
Specifically, the magnification $A$ is a function only of the projected
separation of the lens and the source in units of the Einstein ring, $x$,
which in turn changes with time according to the Pythagorean theorem:
\be
A[x(t)] = {x^2+2\over x\sqrt{x^2+4}},\qquad x(t)=\sqrt{{(t-t_0)^2\over t_e^2}+
\beta^2}.\label{eq:aofxoft}
\ee
Here $t_0$ is the time of maximum magnification and $\beta$ is the impact
parameter in units of the Einstein ring.  Two observers separated by a distance
$d_{\rm sat}$ (projected onto the plane of the sky) will see events with
parameters $(t_0,\beta)$ and $(t_0',\beta')$ respectively.  That is, they
will be separated in the Einstein ring by ${\bf\Delta x}= 
(\omega\Delta t,\Delta \beta)$ where $\omega\equiv t_e^{-1}$.  The magnitude
of $\bf \Delta x$ is related to the physical parameters by simple geometry:
\be
\tilde r_e = {d_{\rm sat}\over \Delta x},\qquad \tilde r_e\equiv 
{\dos\over\dls}r_e,
\ee
where I have now introduced $\tilde r_e$, the Einstein radius projected onto 
the observer plane.  With a parallax satellite, one therefore measures
two parameters $(t_e$ and $\tilde r_e)$ which are combinations of the three
physical quantities $(M$, $v$, and $\dol$).  In addition, the direction
of $\bf\Delta x$ gives the direction of $\bf v$ relative to the known direction
of the satellite.  (The careful reader will have noticed that while 
$\Delta t = t'_0-t_0$ is completely determined, $\Delta \beta = 
\pm(\beta'\pm\beta)$ is determined only up to a four-fold ambiguity.  It
is actually possible to resolve this ambiguity using the slight difference
in $t_e$ induced by the Earth-satellite relative motion.\cite{gouldsatII}
Detailed calculations show this to be feasible for both the LMC~\cite{boutreux}
and the Galactic bulge.\cite{gaudisat})

	What is the value of a parallax satellite?  First, it would give
unambiguous confirmation of the microlensing nature of the events.  Only
two classes of events look significantly different when observed from 1 AU 
apart: microlensing and events within the solar system.  Second, it would
distinguish between events in the Galactic disk, halo, and LMC.  The
quantity $\tilde v\equiv \tilde r_e/t_e$ is very different for these three
populations, being about 50, 300, and 3000 $\kms$ respectively.  Thus, the
populations could be distinguished on an event by event basis.  Third,
$\tilde r_e$ is a function only of the mass and distance.  Hence the individual
(and statistical) mass estimates would be considerably less uncertain than
currently where they are derived from $t_e$ which is a combination of three
physical parameters.  Finally, for some cases it is possible to measure
an additional parameter and so determine the mass, distance, and velocity
separately.

\section{Proper Motions}

	If the lens transits the face of the source, the light curve
deviates from the simple form given by Eq.~\ref{eq:aofxoft} and this permits
one to measure $x_*$, the value of the $x$ when the lens crosses the limb
of the star.\cite{gouldpm}$^{\!-\,}$\cite{wittmao}  Since the angular size
of the source, $\theta_*$ is usually known from its temperature and flux 
(and Sefan's Law), one can therefore determine the angular Einstein radius,
$\theta_e\equiv r_e/\dol$,
\be
\theta_e = {\theta_*\over x_*}.\label{eq:thetaeeval}
\ee
This measurement is called a ``proper motion'' (astronomical jargon for
angular speed) because it immediately yields this quantity, $\mu=\theta_e/t_e$.
As mentioned above, combined parallax and proper motion measurements give
a complete determination of the event parameters.  For example, the mass is
given by
\be
M = {c^2\over 4 G} \tilde r_e\theta_e.\label{mresolve}
\ee
Of course, such transits are rare, but using combined optical/infrared
photometry, one can extend this technique to cases where the lens comes within
two source radii of the source.\cite{gw}  This means that a fraction 
$\sim 2\theta_*/\theta_e$ of events can be measured.  For some classes of
events, such as low-mass objects seen toward the Galactic bulge, this fraction
may be of order 10\%, but for high-mass lenses or LMC events, such 
near-transits are extremely rare.  Hence, many other ideas have been developed
to measure proper motions including interferometric resolution of the two
lensed images~\cite{gouldrev} and lunar occultations~\cite{lunoc} 
(both useful for 
high-mass bulge lenses) as well as spectroscopic measurements of 
spinning~\cite{maoz} and binary~\cite{binsou} source stars (both useful for
LMC events).  Taken together (and in concert with a parallax satellite)
such techniques could give complete solutions for a significant subsample of
events.  This would greatly clarify our understanding of the lenses.

\section{Pixel Lensing of M31}

	In order to observe microlensing, one must monitor stars, generally
millions of them because the optical depth, $\tau$ (the probability that a
given star is being lensed at any give time) is usually $\tau<10^{-6}$.  This
requires dense star fields like the bulge or the LMC.  However, if the fields
get too dense, one cannot resolve the individual stars.  Thus lensing
searches toward galaxies more distant than the LMC would appear impossible
since few if any individual stars are resolved.  Two groups are nevertheless
attempting lensing observations toward M31,\cite{crotts}$^{\!-\,}$\cite{tomcro}
the nearby giant spiral galaxy in Andromeda.  It is always heartening to watch
people attempt the impossible, more inspiring still when they succeed.

	Before discussing why such observations are in fact possible, let me 
focus for a moment on why they are so important.  
First, M31 should have its own halo.  It is highly
inclined so our lines of sight to the far side of the disk pass through much
more of the halo than those toward the near side.  This permits us to study
the structure of the halo along many lines of sight and to test  the
reality of microlensing:  microlensing should be a strong function of position
whereas astrophysical sources like variable stars should be symmetrically
distributed.  By contrast, we have only a few lines of sight along which
to probe the halo of our own Galaxy.  Second, M31 can be used to probe for
exceptionally low mass lenses in our own Galaxy. For observations toward the 
LMC, one is fundamentally restricted to masses $M>10^{-7}M_\odot$.  Smaller
mass lenses have such small angular Einstein rings that they magnify only
a fraction of the surface of a source star.  By contrast, because it is 16
times further away, source stars in M31 are 256 times smaller, so one can
probe to mass scales that are smaller by the same amount.  Third, if dark
matter in the form of lenses is confirmed for the Milky Way, we would like
to begin studying the same stuff elsewhere in the universe, and M31 is a good
place to start.

	Now, why is this impossible project possible?  If a resolved star
is lensed, its flux changes from $F=F_0$ to $F=F_0 A$  The difference is
$\Delta F = F_0(A-1)$.  If the star is unresolved, its light $F_0$ falls on 
some pixel (or more precisely, resolution element) of the detector along with
the light $B$ from many other stars, for a total flux $F_0+B$.  During the
lensing event, the total light grows to $F_0 A+B$, for a difference that is
still $\Delta F = F_0(A-1)$.  Thus, by measuring flux differences of pixels
rather than fluxes of individual stars, one can effectively monitor dozens
of stars in each pixel.  Of course, there are drawbacks to this technique
which come principally from the lower signal-to-noise ratio of the more distant
(and so fainter) stars.  Nevertheless, the main difficulty in carrying out
these projects to date has been a lack of telescope time (due to the 
maddening -- but quite common -- conservativism of telescope time allocation
committees) rather than any fundamental problem with the techniques.  These
projects promise important new results on microlensing within the next few
years and, in addition, are a fantastic way to study previously unobservable
variable stars in external galaxies.

\section{Pixel Lensing of M87}

	Many of the most important astronomical objects (galaxies, globular
clusters, star-forming regions) carry the designation `M' for Messier, an
18th century astronomer.  You might think that he was a far-sighted pioneer who
catalogued objects that would not attract the interest of others for centuries
to come.  Not at all.  Messier was interested in comets which today are
very much an astronomical side show but were the cat's pajamas in the 18th
century.  He often received alleged comet sightings which were then found to
be bogus because the ``comets'' failed to move against the fixed stars.
He therefore constructed a catalogue of this
astronomical garbage so he would not
be distracted from his all-important comet searches.  Eighty-seventh on his
list is now known to be a giant cannibal galaxy at the center of the Virgo
Cluster, the nearest cluster of galaxies to the Milky Way.  

	M87 is an excellent candidate for pixel lensing.  To see why, let's 
think about how Massive Compact Halo Objects (machos) might have formed
in our own Galaxy.  According to the most recent lensing results, the total
macho mass in the Milky Way is $M\sim 2\times 10^{11}\,M_\odot$.  This is
of the same order as the total baryonic material in the visible components
of the Galaxy, its disk and bulge.  Hence, one might imagine that before
the Milky Way had fully collapsed, half of its gas formed into the machos.
The other half collapsed into a proto-disk and proto-bulge which then went
on to form the stars that we see today.  
That would explain the equal amounts of
machos and visible baryons.  Now suppose that this same process went on in
a Milky-Way-like galaxy forming on the outskirts of a cluster.  Half the
gas would still form machos, but before the other half could form stars,
the galaxy would fall through the center of the cluster and be stripped of
its gas by the hot intra-cluster gas.  According to this scenario, there should
then be equal masses of machos and intra-cluster gas in clusters.  The latter
is measured in several clusters to be of order 20\% of the dark matter.
One should then expect an equal amount of machos.  How would one find them?
By pixel lensing of M87, of course!~\cite{gouldmes}  
Since M87 is about 20 times
farther away than M31, the experiment is much more difficult to carry out.
Nevertheless, it would be possible with dedicated observations by HST, with
a time commitment similar to that used on the Hubble Deep Field which has
yielded so many important results.

\section{Planet Searches}

	Microlensing began as a technique to search for dark matter, but
gradually it is being realized that microlensing
has many other potential applications.
The remainder of my contribution is aimed at giving you some flavor of
these possibilities.

	If a planet is orbiting a star and the star is the lens in an
ongoing microlensing event, one could hope to detect the planet through
its perturbations on the lensing light curve.\cite{mp}  
The Einstein ring of the
planet is given by Eq.~\ref{eq:redef}, except with the planet mass $m$ 
replacing the stellar mass $M$.  That is,
\be 
r_p = r_e\sqrt{m\over M}.\label{eq:rpdef}
\ee
Thus, the probability that the planet will perturb the main event at any
given time is only $\sim(r_p/r_e)^2=m/M$ which is too small to be noticed
in standard microlensing searches.  However, the probability that it
will influence the event at {\it sometime} is $\sim(r_p/r_e)=\sqrt{m/M}$.
For Jupiter-mass planets, $m/M\sim 10^{-3}$, this is getting to be an
interesting number.  Moreover, if the planet happens to lie near the Einstein
ring of the lensing star, the effect of the planet is to create a far
reaching ``astigmatism'' in the lens which can dramatically increase the
probability of a detection given intensive monitoring.\cite{gl}  
If a solar-like system lay half way
toward the galactic center, Jupiter would be at $1.3\,r_e$ and its probability
of detection would rise from $\sim 3\%$ to $17\%$.  These effects make
microlensing a potentially powerful tool to search for planets.  If a planet
were detected, one could measure the mass ratio $m/M$ and the
projected planet/star separation in units of the stellar Einstein 
ring.\cite{gl}  In addition, if auxiliary information were available (such
as from a parallax satellite) additional constraints could be placed on the
detected planetary system.

	Two groups are already conducting microlensing searches for 
planets using microlensing ``alert'' events which are recognized in real 
time by the MACHO and OGLE experiments.\cite{planet}~\cite{gman}   The EROS
microlensing experiment has recently begun monitoring the bulge and should
begin producing alerts soon.

	Planets that are substantially smaller than Jupiter are harder
to detect, in part because the probability of an event is lower and in part
because the planet Einstein ring eventually becomes so small that it magnifies
only part of the source star, thus reducing the size of the effect.  
Nevertheless, more ambitious future searches
may be able to detect even Earth-mass planets.\cite{br}$^{\!-\,}$\cite{gdg}

\section{Rotation Speed of Stars}

	One of the ideas that I briefly mentioned for measuring the proper
motion of lensing events was to look for microlensing effects on spinning
stars.\cite{maoz}  The idea is that stellar lines are normally broadening
by the rotation of stars because part of the stellar atmosphere is spinning
toward us and is blue-shifted while the other part is spinning away and so
is redshifted.  During a microlensing event, there is a changing magnification
gradient across the star, which means that either the blue-shifted or 
redshifted side is magnified more than the other.  This causes a shift in 
the centroid of the line, i.e., an apparent shift in the star's radial 
velocity.  The shift is proportional to the rate of stellar rotation and to the
angular size of the star in units of the Einstein ring.  If the first quantity
is measured (from the line broadening) the second can be determined.  This
method is especially useful for rapidly spinning stars (such as A type stars
in the LMC) but it is useless for slow rotators (like K giant type stars in
the bulge) in part because the effect is smaller, but more importantly because
no one knows how fast K giant stars rotate!

	The problem is that the rotation rate is probably an order of magnitude
smaller than the atmospheric turbulence, so the broadening due to rotation
cannot be disentangled from turbulent broadening.  One would very much like
to know how fast K giants rotate in order to learn about the evolution of
angular momentum in stars.  This evolution affects numerous other questions
including rotational mixing and the survival of primeval lithium from the
big bang.

	How can the rotation speed of giants be measured?  By microlensing!
For lensing events where the star transits the source, one can measure
the ratio of the source size to the Einstein ring (see \S\ 4).  
As just stated, the line
shift is proportional to this quantity and to the rotation rate.  Therefore,
if the line shift is also measured then the rotation rate is 
known.\cite{gouldrot}  This is a difficult experiment but a feasible one 
because during a transit event the source star may be magnified by 10 or
20 times, making precision spectroscopy much easier.  In fact, microlensing
is probably the best way to study all aspects of stellar atmospheres,
not just rotation.  Because the lens acts as a giant, ever-changing 
magnifying glass, it in effect resolves the entire surface of the star.

\section{Femtolens Imaging of a Quasar's Black Holes}

	Space constraints prevent me from giving detailed descriptions
of all applications of microlensing.  Before presenting
my final example, let me just mention that microlensing can be used to study 
the star-formation history of the universe,\cite{gouldqso} to measure the
transverse velocities of distant galaxies,\cite{gouldtrans}, to determine
the distribution of binary stars,\cite{mp}~\cite{gdgII} and to search for 
low-mass, high-redshift compact objects~\cite{gouldfem}~\cite{nemgould} such as
primordial black holes or even axion mini-clusters.\cite{kolb}.

	What better way to confirm the hypothesis that quasars are powered by
black holes than to actually image the black hole at the center of a quasar?
To do this, one obviously needs a big telescope since the angular size
of a $10^8\,M_\odot$ black hole at a cosmological distance is only 
$\sim 10^{-9}$ arcsec.  I propose to use a nearby dwarf star as the primary
lens of such a telescope.\cite{femimage}  By a simple calculation, there
should be a dwarf star within about 20 pc of the Sun which lies within a few
arcseconds of a quasar.  Hence, if one were to travel a short distance of
$\sim 40$ AU, the quasar and the dwarf star would be perfectly aligned.  The 
main problem then is to determine what secondary optics must be placed at
focus of the primary lens in order to actually image the black hole.

	The Einstein ring of such a dwarf star is $r_e\sim 10^{-1}$ AU, 
corresponding to $\theta_e\sim 10^{-2}$ arcseconds.  Now most dwarf stars have
binary companions and if the first one you find does not, move on to one
that does.  Typically these companions are at separations $a\sim 10$--100 AU.
As I mentioned above, companions induce an astigmatism on the lens.  However,
for $a\gg r_e$, the astigmatism is very slight and can be completely 
characterized by a shear 
$\gamma\sim (m/M)(r_e/a)^2\sim 10^{-4}$--$10^{-6}$.  The astigmatism is in
the form of a caustic whose size is $\sim 2\gamma$, where here and afterwards
I normalize all sizes to the Einstein ring.  For sources
lying inside the caustic, there are five images.  One of these images lies
near position of the companion and is of no interest here.  The other three
images lie near the Einstein ring.  We maneuver our spacecraft so that the
quasar lies just inside (at a distance $\xi$ from) 
one of the cusps of the caustic. 
Then one of the four images lies on the opposite size of the Einstein ring and
is not highly magnified.  The other three images lie close together on the
same side of the Einstein ring and are highly magnified:  they are stretched
by $\sim \xi^{-1}$ in one direction (along the Einstein ring) and are
contracted by a factor 1/2 in the other direction.  How large a magnification
is possible?  The black hole must fit inside the cusp and this sets a lower
limit $\xi>(4\gamma\rho^2)^{1/3}$ where $\rho\sim 10^{-9}\,$arcsec is the
size of the black hole.  For typical parameters $\xi^{-1}\sim 10^6$.
Unfortunately, because the images are stretched, this permits resolution
of the quasar in only one dimension: all of the information in the other
direction is compressed by a factor of 2 rather than being magnified.  How
can this other information be extracted?

	Consider the three distinct images of a single point.  Each of these
images contains light not only from that point, but from an entire curve
(in fact nearly a straight line) across the source.  The three curves
intersect at exactly one point, the point in question.  In reality, the
resolution elements are finite, so each resolution element contains light
from an entire swath across the source.  The three swaths intersect in
a relatively small region.  Now, let us consider bringing the light from
two of the resolution elements together and then analyzing them with a 
spectrograph.  Since most of the light in each resolution element comes from
non-overlapping regions of the source, it will not interfere.  However, the
light from the intersecting region will appear in both resolution elements and
so will interfere.  By how much?  This depends on the wavelength of the light
and on the relative delay in arrival times of the light along the two paths
represented by the two images.  Rudy Schild discussed his measurement of the
time delay of 1.1 years for the two images of the famous double quasar 
0957+561.\cite{schild}  In the present instance, the time delays are 
$\sim{\cal O}(10^{-15}\rm s)$.  Hence the term femtolensing.  This corresponds
to the wavelength of optical light.  By fourier transformation of the 
interference spectrum, one can determine the amount of light at each position
within the region.  There is addtional information because all three images
can be interfered.  In terms of physical size, the resolution element is
substantially less than the size of the black hole, $\sim 1$ AU.

	The requirements for this project are not trivial.  A set of mirrors
totaling at least 20 square meters must be accurately aligned in a one
dimensional array extending over several hundred meters.  The entire apparatus,
must be sent $\sim 40\,$AU from the Earth and then brought to a velocity
exactly equal to that of the dwarf star (so that it remains aligned with
the star-quasar line of sight).  It will even be necessary to correct this
velocity every few hours to compensate for the gravitational effects of the
Sun and the accelerated motion due to the star's companion.  But microlensing
is an incredibly powerful tool and we should be ambitious in deploying it.

\section*{Acknowledgments}
This work was supported in part by NSF grant AST 9420746.

\end{document}